\documentclass[10pt]{article}
\usepackage{graphicx}
\usepackage{lineno}
\usepackage{subfigure}

\def\Title#1{\begin{center} {\Large #1 } \end{center}}
\def\Author#1{\begin{center}{ \sc #1} \end{center}}
\def\Address#1{\begin{center}{ \it #1} \end{center}}

\newcommand\pubblock{\rightline{\begin{tabular}{l} Proceedings of the Second Annual LHCP\\ \pubnumber\\
         \pubdate  \end{tabular}}}

\newenvironment{Abstract}{\begin{quotation} \begin{center} 
             \large ABSTRACT \end{center}\bigskip 
      \begin{center}\begin{large}}{\end{large}\end{center} \end{quotation}}

\newenvironment{Presented}{\begin{quotation} \begin{center} 
             PRESENTED AT\end{center}\bigskip 
      \begin{center}\begin{large}}{\end{large}\end{center} \end{quotation}}

\def\beq{\begin{equation}}
\def\eeq#1{\label{#1}\end{equation}}
\def\eeqn{\end{equation}}

\def\beqa{\begin{eqnarray}}
\def\eeqa#1{\label{#1}\end{eqnarray}}
\def\eeqan{\end{eqnarray}}

\let\bar=\overbar

\def\Dslash{\not{\hbox{\kern-4pt $D$}}}
\def\dslash{\not{\hbox{\kern-2pt $\del$}}}

\def\msb{{\bar{\ssstyle M \kern -1pt S}}}

\usepackage{color}
\usepackage{cite}

\newcommand\pubnumber{ ATL-PHYS-PROC-2014-097 }

\newcommand\pubdate{August 12, 2014}

\def\affiliation{
On behalf of the ATLAS Experiment, \\
Department of Physics,\\
University of Washington, 
Box 351560, Seattle WA 98195-1560, USA}

\begin{document}


\large
\begin{titlepage}
\pubblock

\vfill
\Title{ Beyond Standard Model Higgs boson physics with the ATLAS experiment
at the LHC }
\vfill

\Author{ Nikolaos Rompotis  }
\Address{\affiliation}
\vfill
\begin{Abstract}
The search for evidence of beyond Standard Model Higgs bosons is an
integral part of the Higgs boson studies
at the LHC. This article reviews recent beyond Standard Model Higgs boson
searches using Run I LHC proton-proton collision 
data  recorded by the ATLAS detector. In particular, searches
for Higgs boson cascades, double Higgs boson production, scalar particles
decaying to $\gamma\gamma$ pairs, flavor changing neutral currents
involving Higgs bosons, and  
 Higgs bosons decaying to invisible particles are discussed.
No significant  deviations from the background expectations are found 
and corresponding constraints on physics beyond the 
Standard Model are obtained.
\end{Abstract}
\vfill

\begin{Presented}
The Second Annual Conference\\
 on Large Hadron Collider Physics \\
Columbia University, New York, U.S.A \\ 
June 2-7, 2014
\end{Presented}
\vfill
\end{titlepage}
\def\thefootnote{\fnsymbol{footnote}}
\setcounter{footnote}{0}
%

\normalsize 


\section{Introduction}

The recent observation of a new particle with mass 125--126~GeV at the 
LHC \cite{Aad:2012tfa,Chatrchyan:2012ufa} opens 
a new era for particle physics. 
The detailed study of the production and decay modes of the 
new particle provides invaluable input to answer the question 
of whether this is indeed the long-sought Standard Model (SM) 
Higgs boson \cite{Englert:1964et,Higgs:1964ia,Higgs:1964pj,Guralnik:1964eu,Higgs:1966ev,Kibble:1967sv}.
The first measurements indicate that the new particle is indeed 
compatible with the SM Higgs boson, see e.g., \cite{Aad:2013wqa}. 
Nevertheless, many more measurements and data will be needed
to extract reliable conclusions. This task is further complicated 
by the fact that many beyond SM physics scenarios include a 
SM-like Higgs boson, which is part of an extended scalar sector.
In that case, searches for beyond SM Higgs bosons are very 
interesting, since they provide direct information on a 
possibly extended scalar sector, and hence they are complementary
to the precise measurements of the properties of the new particle.

This article  reviews some of the recent searches 
for beyond SM Higgs bosons or exotic properties of the recently discovered
Higgs boson  using 
proton-proton collision data at 7 and 8 TeV center-of-mass energies 
as recorded by the ATLAS detector \cite{Aad:2008zzm}.

\section{Higgs Bosons in beyond SM physics scenarios}

The scalar sector of the SM consists of a complex Higgs doublet, 
which after the electroweak symmetry breaking  leaves a 
single scalar boson in the theory. Possible extensions of the 
scalar sector are restricted due to the fact 
that, in general, they include unacceptably large 
corrections to the well-measured
quantity $\rho\equiv m_W^2/(m_Z^2\cos^2\theta)$, where 
$\theta$ is the weak mixing angle \cite{Gunion:1989we}.
The addition of doublets or singlets is such that leading
order corrections to $\rho$ vanish and, hence they can be 
compatible with precision electroweak tests \cite{Baak:2012kk}.

The introduction of one additional
Higgs doublet defines a class of models, which are collectively known
as 2 Higgs doublet models (2HDM) \cite{Lee:1973iz}.
These models introduce five Higgs bosons, three of which are neutral
and two of which are charged.
The Minimal Supersymmetric Standard Model (MSSM)
\cite{Fayet:1976et,Fayet:1977yc,Farrar:1978xj,Fayet:1979sa},
which is a very popular realization of supersymmetry,
is a particular case of a 2HDM. The MSSM, as well as other
2HDMs, are compatible with existing measurements, including
the recently discovered Higgs boson 
\cite{Bechtle:2012jw,Branco:2011iw}.

Apart from the introduction of new particles, extensions of the
SM scalar sector may affect the properties of the SM-like Higgs boson
discovered at the LHC.
Enhancement of rare decays, completely new decays 
and production mechanisms, 
and different couplings with respect to the SM expectations are possible
and may indicate connections to other puzzles, such as dark matter
(e.g., see Ref.~\cite{Curtin:2013fra} and references therein).

In the following, only a few recent ATLAS searches are described.
As a convention, the symbol $h$ will refer to the newly
discovered Higgs boson with mass $\sim 125$~GeV.

\section{Some recent searches for beyond SM Higgs bosons with ATLAS}

\begin{figure}[htb]
\centering
\subfigure[]{\label{fig:cascade-1}
\includegraphics[width=0.45\textwidth]{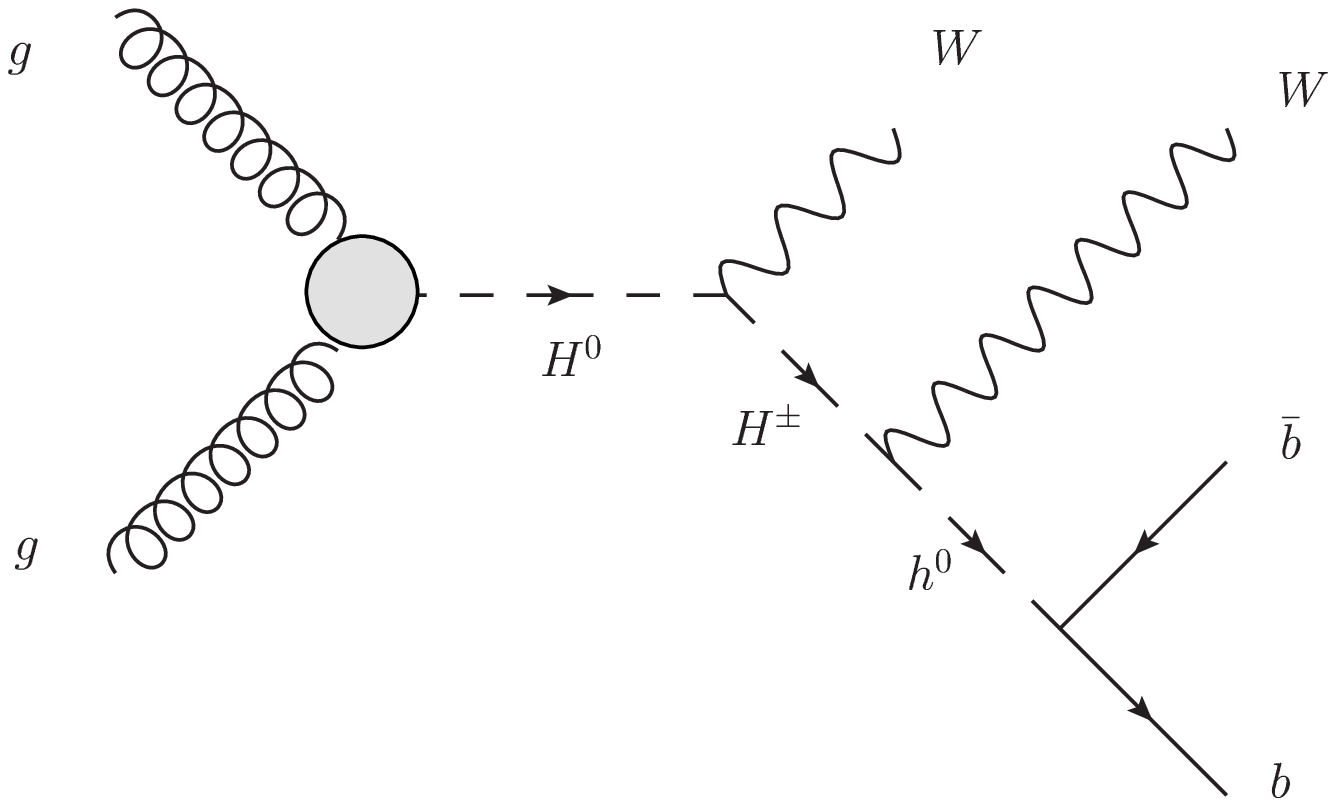}}
\hspace{1.0cm}
\subfigure[]{\label{fig:cascade-2}
\includegraphics[width=0.45\textwidth]{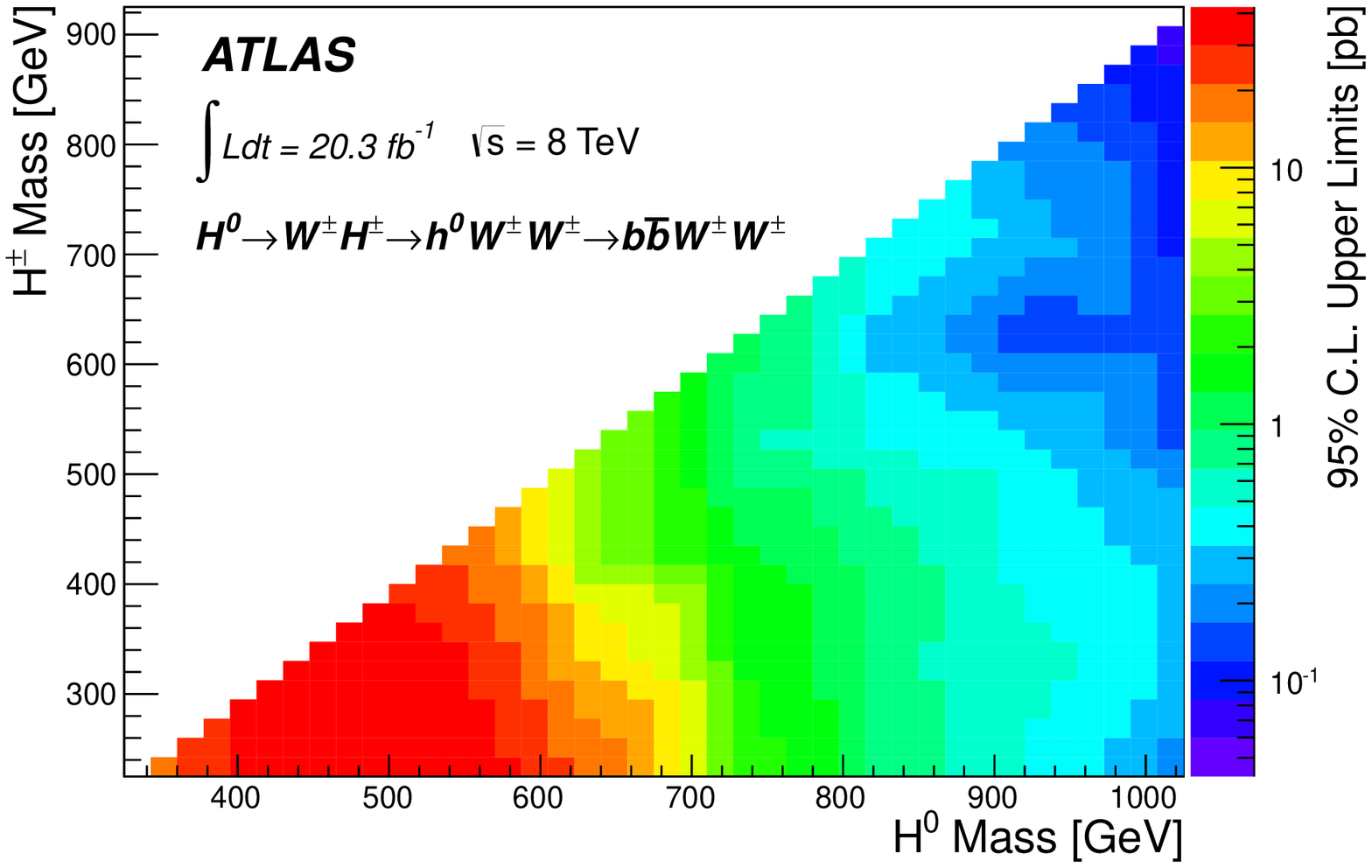}}
\caption{The topology of a Higgs cascade search, described
in Ref.~\cite{Aad:2013dza}, is shown in (a). The observed upper limit
for this process is shown in (b).}
\label{fig:cascade}
\end{figure}

The existence of an extended scalar sector with more than one 
Higgs bosons opens the possibility of cascade decays in which 
heavier Higgs bosons decay into lighter ones. In this context, 
a Higgs boson cascade has been
looked for in Ref.~\cite{Aad:2013dza} using 20.3~fb$^{-1}$ of
 8 TeV proton-proton collision data. A heavy Higgs boson
is produced via gluon-gluon fusion and initiates a cascade decay 
that includes a charged scalar, $H^{\pm}$, a light neutral Higgs 
boson, $h$,  and
$W$ bosons: $H \rightarrow W^{-} H^{+} \rightarrow  W^{-} W^{+} h$,
see Fig.~\ref{fig:cascade-1}.
The light neutral Higgs boson decays to $b\bar{b}$ and it 
is assumed to have mass 125~GeV. 
The first of the $W$ bosons decays leptonically and provides a way
to trigger the events whereas the second one decays hadronically
maximizing the branching ratio.
The final state shares the same topology with  $t\bar{t}$
events with similar $W$ boson
decays, which is the main background process for the search.
A multivariate discriminant is employed to exploit the kinematic 
differences
between signal and $t\bar{t}$ events. Upper limits on the production
cross section of heavy neutral Higgs boson, $H$, times the
branching ratio 
BR$(H\rightarrow W^{\mp}H^{\pm} \rightarrow W^{\mp}W^{\pm}h \rightarrow  W^{\mp}W^{\pm}b\bar{b})$
 are derived as a function of its mass
and the charged scalar mass, as shown in Fig.~\ref{fig:cascade-2}.

\begin{figure}[htb]
\centering
\includegraphics[width=0.45\textwidth]{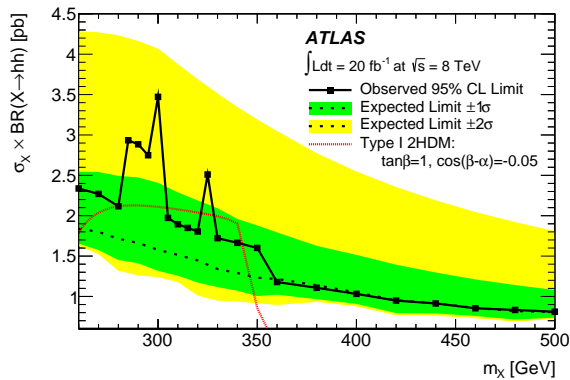}
\caption{Upper limits from the search for a scalar particle decaying to
$hh$ in the $b\bar{b}\gamma\gamma$ channel from Ref~\cite{Aad:2014yja}.}
\label{fig:figure_bbgg}
\end{figure}

The pair production of  Higgs bosons 
is very interesting and can be
enhanced due to the resonant decay of a heavy CP-even Higgs boson
($H\rightarrow hh$), e.g., in a 2HDM, or even in a non-resonant
way, e.g., in composite Higgs boson models~\cite{Gillioz:2012se}.
The analysis described in Ref.~\cite{Aad:2014yja} uses 20.3~fb$^{-1}$
of 8 TeV proton-proton collision data to look for $hh$ production
in the $h\rightarrow b\bar{b}$ and $h\rightarrow \gamma\gamma$ channel.
Both resonant and non-resonant anomalous production of $hh$ pairs
is searched for. Upper limits on the production of a narrow-width
heavy scalar boson decaying to $hh$ as a function of its mass
are shown in Fig.~\ref{fig:figure_bbgg}. The cross section for
non-resonant $hh$ production is constrained to be less than
$2.2$~pb at 95\% confidence level. As a reminder, the SM
$hh$ production cross section is  $\sim 10$~fb, i.e., about two 
orders of magnitude smaller than the sensitivity of this search.

\begin{figure}[htb]
\centering
\includegraphics[width=0.95\textwidth]{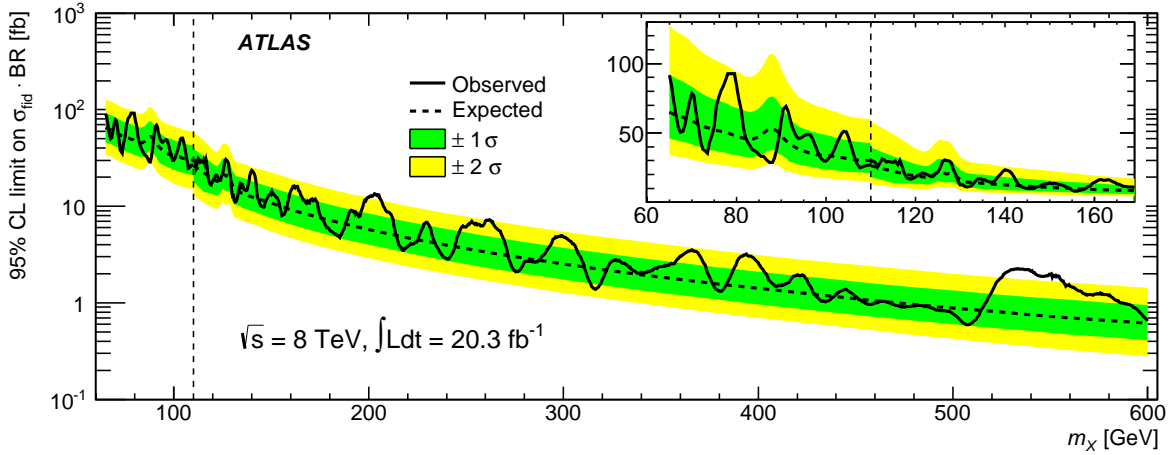}
\caption{Upper limits at 95\% confidence level of a search for a
scalar particle decaying to a $\gamma\gamma$ pair from 
Ref.~\cite{Aad:2014ioa}. }
\label{fig:figure_gammagamma_2}
\end{figure}

A search for a scalar particle that decays to a $\gamma\gamma$ pair
and is in the mass range 65--600~GeV is described 
in Ref.~\cite{Aad:2014ioa} and uses 20.3~fb$^{-1}$ of 
8 TeV proton-proton collision data.
This analysis considers the SM-like Higgs boson at $\sim 125$~GeV
as a background to the search. Analytical functions are employed to 
describe the shape of the $\gamma\gamma$ invariant mass similar to
those used in the SM Higgs boson search in the $\gamma\gamma$ channel
\cite{Aad:2012tfa}. The $\gamma\gamma$ mass range of this search is
constrained by trigger and background estimation requirements 
from the low mass side and by the data statistics 
in the high mass side. Upper limits
on the fiducial cross section times the branching ratio 
$H\rightarrow\gamma\gamma$ are quoted, 
see Fig.~\ref{fig:figure_gammagamma_2}.

\begin{figure}[h]
\centering
\subfigure[]{ \label{fig:figure_htc-1}
\includegraphics[width=0.45\textwidth]{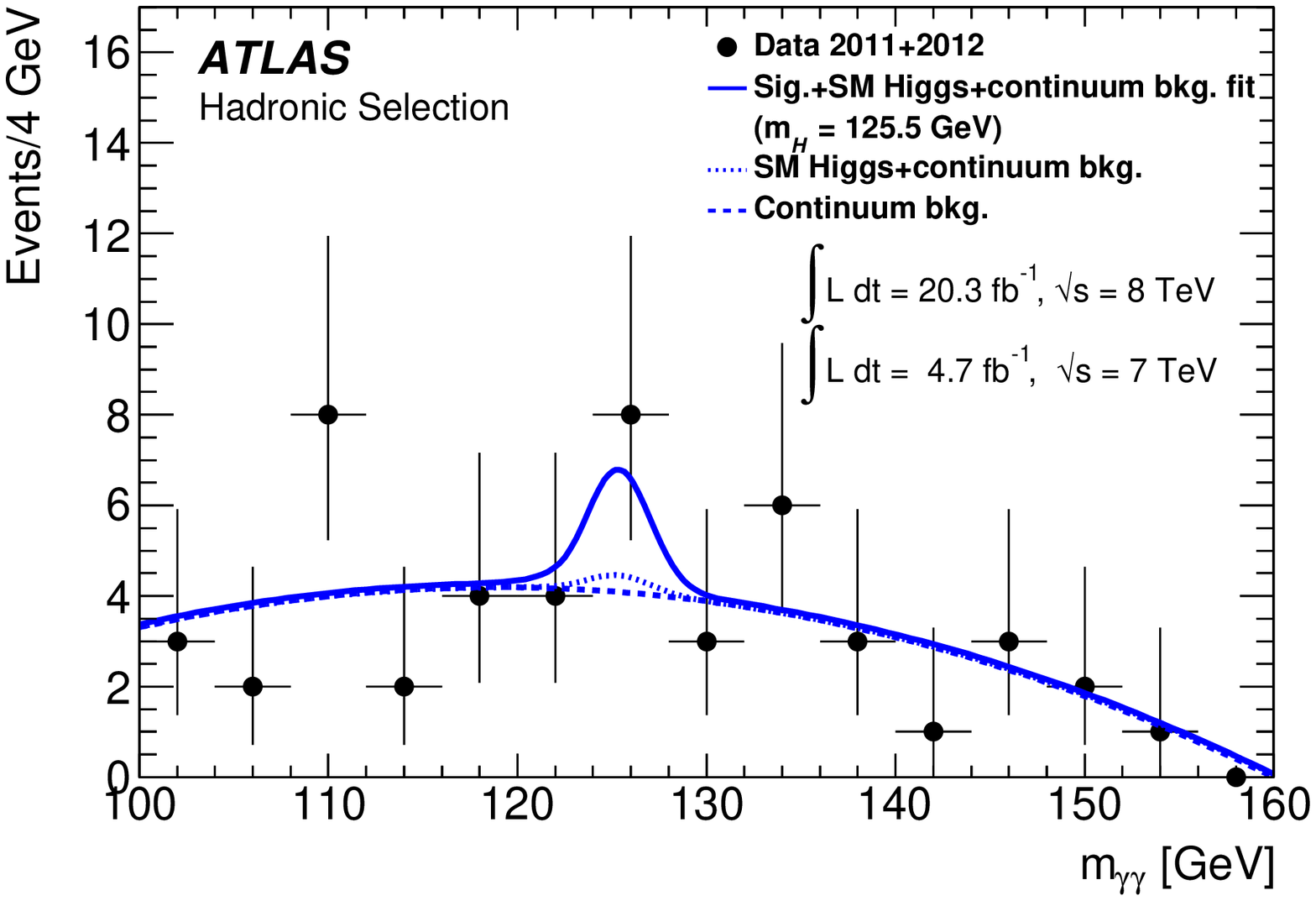} }
\subfigure[]{ \label{fig:figure_htc-2}
\includegraphics[width=0.45\textwidth]{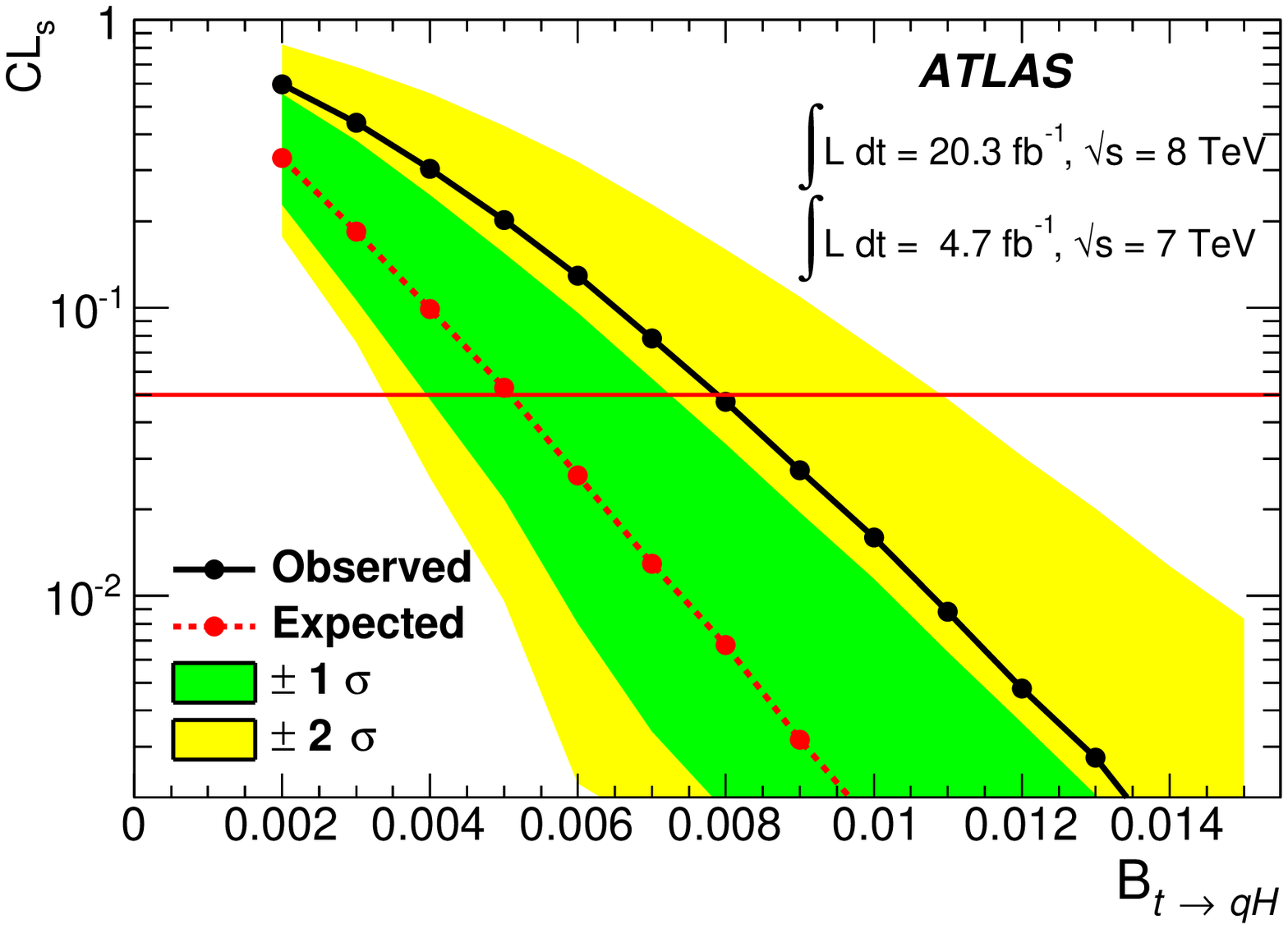} }
\caption{The distribution of the $\gamma\gamma$ mass for
$t\bar{t} \rightarrow hq Wb$ candidates in which the $W$ boson 
decays hadronically is shown in (a). Observed and expected
CLs for the $t\rightarrow hq$ search is shown in (b). Both
figures are from Ref.~\cite{Aad:2014dya}.}
\label{fig:figure_htc}
\end{figure}

Flavor changing neutral currents (FCNC) are in general enhanced 
in extended Higgs sectors. In several occasions, models can be 
built that evade the tight constraints on FCNC by flavor physics, 
like the type-III 2HDM \cite{Branco:2011iw}. 
In those cases, the $thc$ coupling is predicted to be sizeable
and close to the LHC sensitivity \cite{Chen:2013qta}.
A search for top quark decaying to $hq$, 
where $q$ denotes some light quark, 
is reported in Ref.~\cite{Aad:2014dya} and uses 4.7~fb$^{-1}$ of
7 TeV and 20.3~fb$^{-1}$ of 8 TeV proton-proton collision data.
This analysis looks for $t\bar{t}$ production in which one top
decays to $hq$ and the other semileptonically or in a 
fully hydronic way.
An example of the $\gamma\gamma$ invariant mass after all selection
requirements for the case where one of the tops decays in a fully
hydronic way is shown in Fig.~\ref{fig:figure_htc-1}.
The observed (expected) upper limit at 95\% confidence level
for the flavor changing branching
ratio is BR($t\rightarrow hq) < 0.79 (0.51)$\%, 
see Fig.~\ref{fig:figure_htc-2}.  

\begin{figure}[htb]
\centering
\subfigure[]{ \label{fig:figure_invisible-1}
\includegraphics[width=0.45\textwidth]{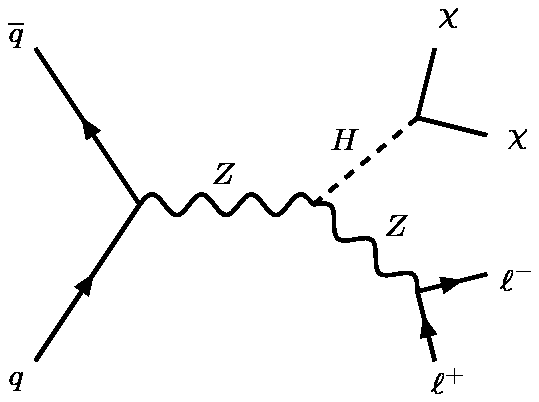}}
\subfigure[]{ \label{fig:figure_invisible-11}
\includegraphics[width=0.45\textwidth]{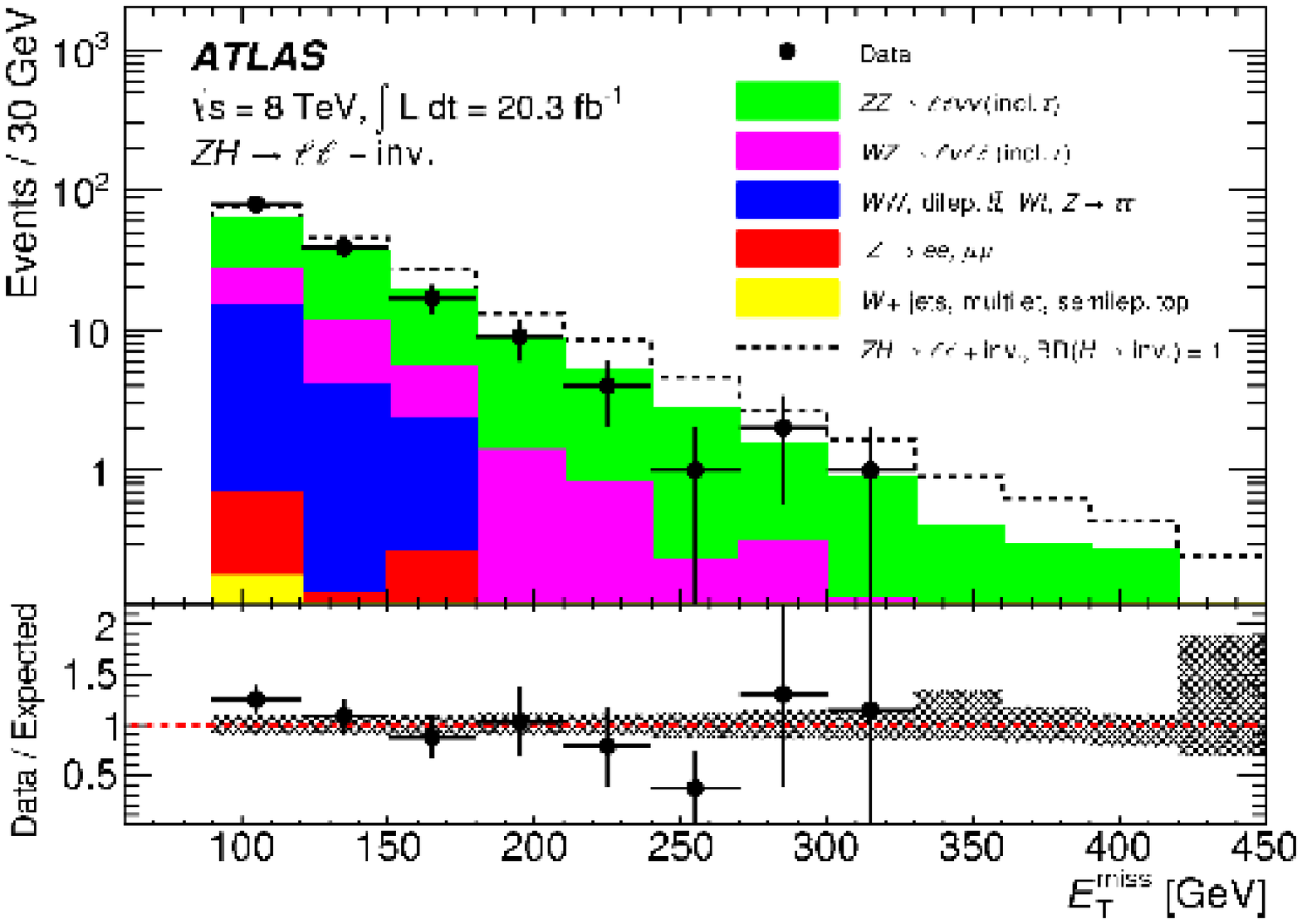}}
\subfigure[]{ \label{fig:figure_invisible-2}
\includegraphics[width=0.45\textwidth]{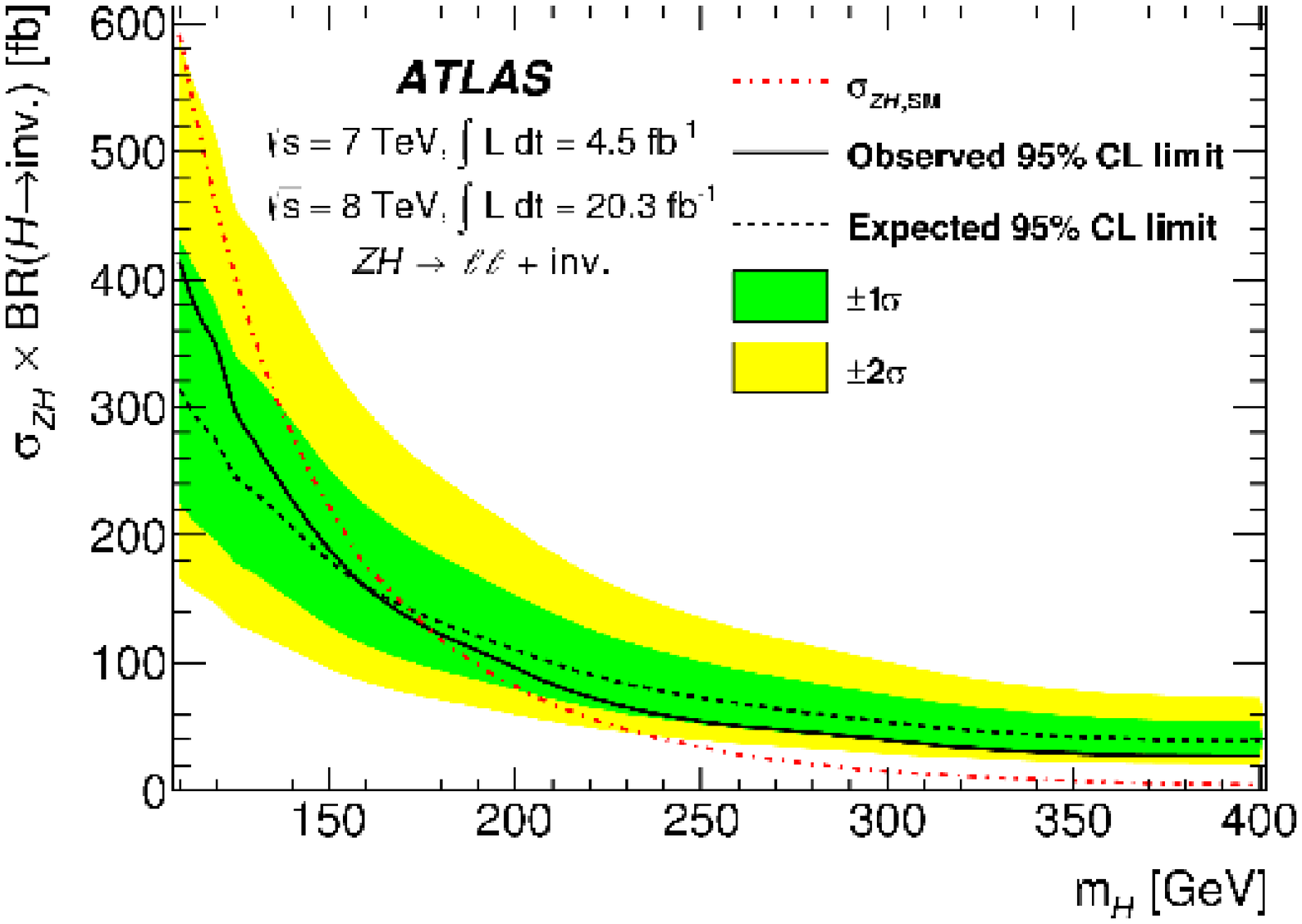}}
\caption{The topology of the search for invisible Higgs boson
decays described in Ref.~\cite{Aad:2014fia} is shown in (a).
The missing transverse momentum after the full selection
and the upper limits on this process are shown in (b) and (c),
respectively.}
\label{fig:figure_invisible}
\end{figure}

A search for invisible decays of the Higgs boson discovered
at the LHC can provide access to dark matter through the
Higgs portal \cite{Patt:2006fw}. 
A search reported in Ref.~\cite{Aad:2014fia} 
uses  4.5~fb$^{-1}$ of
7~TeV and 20.3~fb$^{-1}$ of 8~TeV proton-proton collision data 
to look for the associated production
of a Higgs boson with a $Z$ boson as shown pictorially 
in Fig.~\ref{fig:figure_invisible-1}. The $Z$ boson decays
to an electron or a muon pair and the Higgs boson is assumed to decay
to invisible particles. 
The final event sample is dominated by $Z$+jet events and the
missing transverse momentum distribution is examined for
discrepancies with respect to the SM prediction, see 
Fig.~\ref{fig:figure_invisible-11}
Upper limits for the  production of a Higgs boson in association
with a $Z$ boson, assuming that the Higgs boson decays always to
invisible particles, 
are shown in Fig.~\ref{fig:figure_invisible-2}.
This result is combined with indirect constraints on invisible
Higgs boson decays from Higgs boson coupling measurements
\cite{ATLAS-CONF-2014-010}. 
The observed (expected) combined constraint 
for the 125~GeV Higgs boson on the 
invisible branching ratio is 
BR$(h\rightarrow \mathrm{invisible}) < 37 (39)$\% at 95\% confidence level.

\section{Conclusions}

The search for beyond SM Higgs bosons is highly motivated and has just 
started having sensitivity to realistic scenarios. The discovery 
of a Higgs boson with mass around 125 GeV has opened new 
possibilities for searches, especially those that include 
the new particle in the final state.
More data are needed in order to explore the possibility 
of an extended
scalar sector and, hence the community is looking forward to the 
new LHC run.



\end{document}